\documentclass[prd,11pt]{revtex4-1}

\usepackage{bm}
\usepackage{amsmath}    
\usepackage{graphicx}   
\usepackage{verbatim}   
\usepackage{color}      
\usepackage{subfigure}  
\usepackage{hyperref}   
\definecolor{amaranth}{rgb}{0.9, 0.17, 0.31}
\definecolor{purple(munsell)}{rgb}{0.62, 0.0, 0.77}
\definecolor{americanrose}{rgb}{1.0, 0.01, 0.24}
\definecolor{palatinateblue}{rgb}{0.15, 0.23, 0.89}
\definecolor{royalblue(web)}{rgb}{0.25, 0.41, 0.88}
\definecolor{hanpurple}{rgb}{0.32, 0.09, 0.98}
\definecolor{beaublue}{rgb}{0.74, 0.83, 0.9}
\definecolor{carminered}{rgb}{1.0, 0.0, 0.22}
\definecolor{brightpink}{rgb}{1.0, 0.0, 0.5}
\definecolor{vividviolet}{rgb}{0.62, 0.0, 1.0}

\hypersetup{ linktoc=all,
    colorlinks, linkcolor={palatinateblue},
    citecolor={brightpink}, urlcolor={amaranth}}

\newcommand{\be}{\begin{equation}}
\newcommand{\ee}{\end{equation}}
\newcommand{\bs}{\begin{split}} 
\newcommand{\bea}{\begin{eqnarray}}
\newcommand{\eea}{\end{eqnarray}}

\begin{document}

\title{ Dirac Oscillator in DSR: A Comparative Study of  Magueijo-Smolin and Amelino-Camelia Models}

\author{Nosratollah Jafari$^1$ \footnote{\email{  nosrat.jafari@fai.kz (Corr. Author)}} ,~ 
Abdelmalek Boumali$^2$  \footnote{\email{abdelmalek.boumali@univ-tebessa.dz; boumali.abdelmalek@gmail.com}}
\smallskip 
\\
$^1$Fesenkov Astrophysical Institute, 050020, Almaty, Kazakhstan
\\
$^1$ Al-Farabi Kazakh National University, Al-Farabi av. 71, 050040 Almaty, Kazakhstan
\\
 $^1$ Center for Theoretical Physics, Khazar University, 41 Mehseti Street, Baku, AZ1096, Azerbaijan
\\
$^2$ Laboratory of theoretical and applied Physics Echahid Cheikh Larbi Tebessi University, Algeria}

\begin{abstract}

This paper investigates the energy spectrum of the Dirac oscillator within the framework of Doubly Special Relativity (DSR), focusing on two prominent models: the Magueijo–Smolin (MS)  and Amelino-Camelia models.   We derive the modified Dirac equations in both MS and Amelino-Camelia  DSR models under the approximation of $O(E^2/k^2)$ for a single particle and examine the resulting energy spectra.   The study reveals significant corrections to the standard relativistic Dirac oscillator spectrum due to the Planck-scale deformation parameter $k$, which introduces distinct deviations depending on the DSR model employed.   For the MS model, we observe non-uniform shifts in both positive and negative energy branches at small $k$, with the spectrum gradually flattening towards the canonical result as $k$ increases.   In the Amelino-Camelia model, the energy levels show larger deviations at lower values of $k$, and these anomalies diminish more slowly compared to the MS model.   The results provide valuable insights into the impact of quantum gravity effects on quantum systems, with potential applications in high-precision spectroscopic or astrophysical observations at energies near the Planck scale.   Furthermore, the comparative analysis of these two DSR models highlights the robustness of Planck-scale predictions and guides future experimental efforts to detect quantum-gravity signatures.

\vspace{7.5cm}


\end{abstract}

\maketitle

\tableofcontents

\section{Introduction}

The relativistic extension of the harmonic oscillator ranks among the few quantum systems that admit exact solutions. In its Dirac form, one replaces the canonical momentum $\vec p$ by $\vec p - i m \beta \omega \vec r$ in the Dirac equation—a prescription first proposed by Ito et al.\cite{ito1967}. Moshinsky and Szczepaniak later coined the term “Dirac oscillator” (DO), showing that in the nonrelativistic limit it recovers a harmonic oscillator augmented by a strong spin–orbit coupling\cite{moshinsky1989}. Physically, the DO interaction can be interpreted as the coupling of an anomalous magnetic moment to a linearly growing electric field\cite{moreno1989,martinez1992}, and the corresponding electromagnetic potential was explicitly derived by Benitez et al.\cite{Benetez1990}. The DO’s exact solvability and its applications—spanning nuclear structure, particle phenomenology, and quantum optics—have sustained extensive theoretical interest\cite{quesne1990,quesne2005,Boumaliejtp2015,Quimbay2013}. Its first proposed experimental realization in a one-dimensional microwave resonator was presented by Franco-Villafane et al.\cite{Franco-Villafane2013}, cementing the DO as a vital bridge between relativistic quantum mechanics and laboratory models\cite{boumali2020,guvendi2021,pacheco2003,pacheco2014,Guvendi:2025det,Guvendi:2024gds}.  

DSR generalizes Einstein’s framework by postulating an additional invariant scale, the Planck energy $ k = \sqrt{\hbar c^5/G}\approx10^{19}\,\mathrm{GeV}$, alongside the speed of light $c$. While conventional special relativity preserves only $c$-invariance, DSR deforms the energy–momentum relations to encode potential quantum-gravitational effects at ultra-high energies. Two seminal DSR models are the Amelino-Camelia proposal\cite{Amelino-Camelia:2000stu,Amelino-Camelia:2002uql} and the Magueijo–Smolin (MS) construction\cite{Magueijo:2001cr}, both of which maintain observer independence of $c$ and $k$\cite{JafariPhysRevD}. These frameworks offer phenomenological pathways to probe Planck-scale physics, and relativistic oscillators like the DO serve as ideal testbeds for exploring the interplay between quantum mechanics, special relativity, and gravity-induced deformations\cite{JafariPhysLettB2020,JafariPhysLettB2024,JafariPhysLettB2025,JafariPhysRevD,GuvendiDSRPhysLettB2024,Coraddu:2009sb}.  

The primary objectives of this paper are:   (i) to determine how the DSR deformation parameter influences the energy spectrum of a 1D Dirac oscillator, with emphasis on the bifurcation of particle- and antiparticle-like branches;   (ii) to contrast the spectral modifications predicted by the Amelino-Camelia and MS models;   (iii) to identify the regimes in which the spectrum remains real and to elucidate the approach to the classical limit as the deformation scale becomes large.  

The paper is organized as follows: Section \ref{SolutionMS} presents the analogous derivation for the MS model and analyzing its spectral features.  Section \ref{SolutionAC} derives and solves the 1D DO  modified by the Amelino-Camelia DSR framework, and finds the resulting spectra. Finally, Section \ref{Conclusion} summarizes the main findings by comparing  the MS and the Amelino-Camelia  cases and discusses their implications for quantum-gravity phenomenology.

\section{Solutions of the Dirac oscillator in MS DSR} \label{SolutionMS}

The modified Dirac equation in MS DSR, in the \( O(E^2 /  k^{2}) \) approximation for a single particle, can be expressed as follows \cite{Coraddu:2009sb}:
\begin{equation}
\left[ i \gamma^\mu \frac{\partial}{\partial x^\mu} -  m \left(1- \frac{i}{k} \frac{\partial}{\partial t} \right)  \right]\tilde{\psi}   =0,\end{equation}
or 
\be \label{MDiracAlpha}
\left\{\cdot \alpha_x \cdot\left(p_x-i m \omega \beta \cdot x\right)+\beta M\right\} \psi_D=E \psi_D,
\ee
where $M=m \left(1- \frac{i}{k} \frac{\partial}{\partial t} \right)$
with $\psi_D=\left(\psi_1 \psi_2\right)^T, \alpha_x=\sigma_x$ and $\beta=\sigma_z$.
From Eq. (\ref{MDiracAlpha}), we get a set of coupled equations as follows 

\bea \label{EquationforEtwo}
\begin{aligned}
& \left(E-M\right) \psi_1=\left(p_x+i m \omega x\right) \psi_2, \\
& \left(E+M\right) \psi_2=\left(p_x-i m \omega x\right) \psi_1 .
\end{aligned}
\eea
Using Eq. (\ref{EquationforEtwo}), we have
\be\label{EqPsi2}
\psi_2(x)=\frac{\left(p_x-i m \omega x\right)}{E+M} \psi_1(x) .
\ee
Putting Eq. (\ref{EqPsi2}) into Eq. (\ref{MDiracAlpha}), we get

\be \label{EqE^2px}
\left[\left(p_x+i m \omega x\right)\left(p_x-i m \omega x\right)-E^2+M^2\right] \psi_1(x)=0,
\ee
or
\be \label{EqE^2M^2}
\left(\frac{p_x^2}{2 m}+\frac{m \omega^2}{2} x^2\right) \psi_1(x)=\left(\frac{\omega m +E^2-M^2 }{2 m}\right) \psi_1(x) \equiv \tilde{E} \psi_1 .
\ee

Equation (\ref{EqE^2M^2}) coincides with the familiar one‐dimensional harmonic‐oscillator equation, whose energy eigenvalues are well known and given by

\be
\left(\frac{\omega m +E^2-m^2(1-\frac{E}{k})^2 }{2 m}\right)=\omega \left( n+1/2 \right)
\ee
The resulting eigenvalues are
\be
E=-\frac{m^{2}k}{k^{2}-m^{2}}\pm\frac{km\sqrt{k^{2}+2k^{2}n\frac{\omega}{m}-2mn\omega}}{k^{2}-m^{2}}
\ee

\begin{figure*}
    \centering
    \includegraphics[scale=0.65]{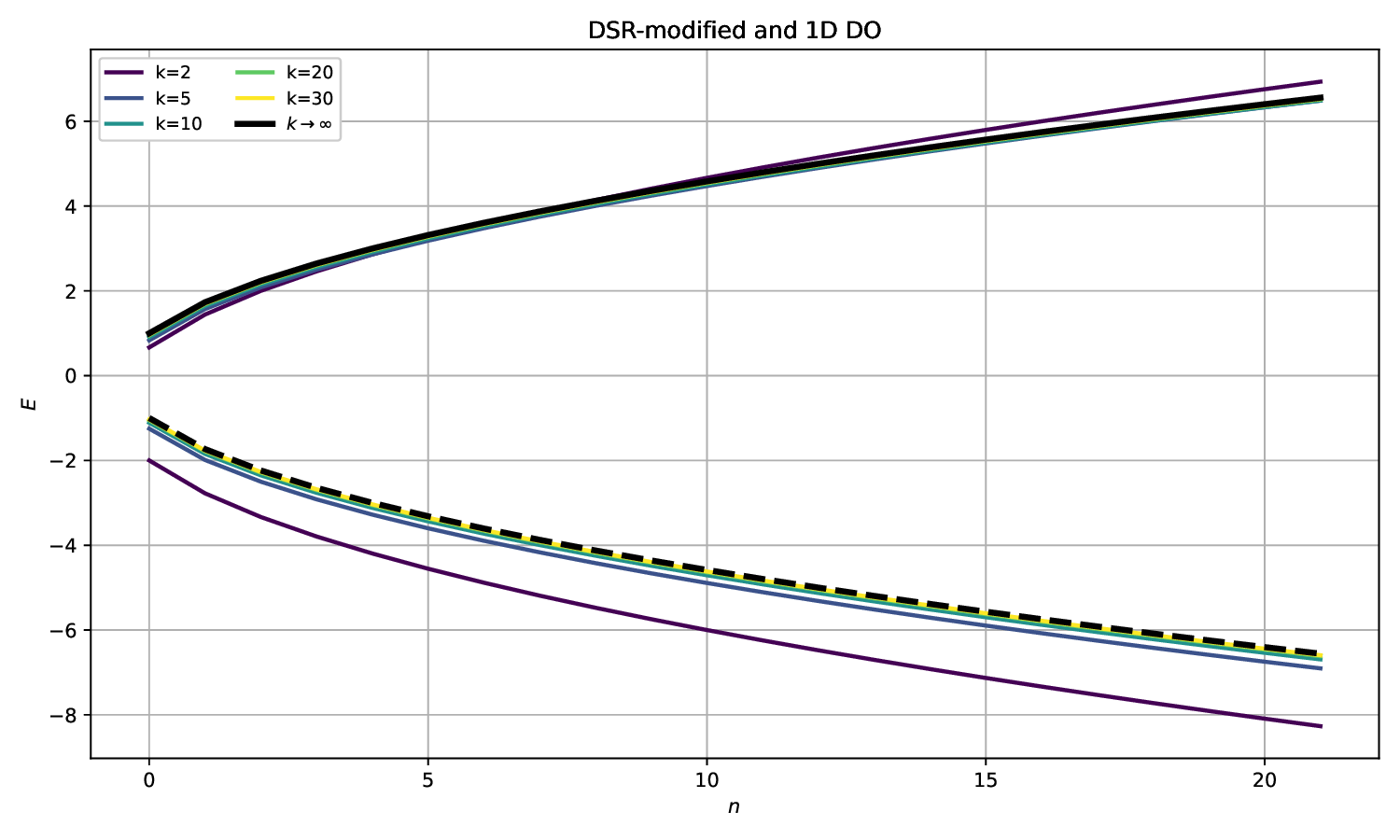}
    \caption{Plot of Energy Levels for the 1D Dirac Oscillator in MS DSR for Various Values of $k$}
    \label{fig1}
\end{figure*}

The energy spectrum in this form is illustrated in Figure. \ref{fig1}.   This figure presents a detailed representation of the spectrum of the 1d DO in MS DSR , showcasing how the energy levels behave under the specific conditions considered.   It highlights the relationship between energy and the quantum number \$n\$, as well as the influence of the parameter $k$ on the spectrum's structure.   By visually capturing these elements, Figure 1 offers a comprehensive view of the variation of the energy spectrum, allowing a deeper understanding of how the different parameters affect the overall energy profile in the context under study.
In the limit when $ k \rightarrow \infty $ we recover the standard formula of 1D DO.
\be
E=\pm m\sqrt{1+2n\frac{\omega}{m}}
\ee

\section{Solutions of the Dirac oscillator  in Amelino-Camelia DSR} \label{SolutionAC}

The modified Dirac equation in Amelino-Camelia's DSR, in the \( O(E^2 /  k^{2}) \) approximation for a single particle, can be expressed as follows \cite{Jafari:2024bph},
\begin{equation}
\left[ i \gamma^0  \frac{\partial}{\partial t} + i \gamma^i \frac{\partial}{\partial x^i} \left(1 + \frac{i}{2k} \frac{\partial}{\partial t} \right) - m \right] \tilde{\psi} = 0. \label{MDE}
\end{equation}
or 
\be \label{ACDiracPx}
\left\{\alpha_x \cdot\left(p_x-i m \omega \beta \cdot x\right)\right\} \psi_D=\left( \frac{E}{\kappa} - \frac{m}{\kappa} \beta \right) \psi_D,
\ee
with $\kappa = 1+\frac{E}{2k}$.

From Eq. (\ref{ACDiracPx}), we get a set of coupled equations as follows 

\bea \label{ACDiractwo}
\begin{aligned}
& \left(\frac{E-m}{\kappa} \right) \psi_1=\left(p_x+i m \omega x\right) \psi_2, \\
& \left(\frac{E+m}{\kappa} \right) \psi_2=\left(p_x-i m \omega x\right) \psi_1 .
\end{aligned}
\eea
Using Eq. (\ref{ACDiractwo}), we have
\be \label{ACPsi}
\psi_2(x)=\frac{\kappa \left(p_x-i m \omega x\right)}{E+m} \psi_1(x) .
\ee
Putting Eq. (\ref{ACPsi}) into Eq. (\ref{ACDiracPx}), we get

\be 
\left[\left(p_x+i m \omega x\right)\left(p_x-i m \omega x\right)+\frac{m^2-E^2}{\kappa^2}\right] \psi_1(x)=0,
\ee
or
\be 
\left(\frac{p_x^2}{2 m}+\frac{m \omega^2}{2} x^2\right) \psi_1(x)=\left(\frac{\omega}{2}+\frac{E^2-m^2}{2m \kappa^2} \right) \psi_1(x) \equiv \tilde{E} \psi_1 .
\ee
The eigen solutions are
\be
\frac{\omega}{2}+\frac{E^2-m^2}{2m \left( 1+\frac{E}{2k}\right)^2}=\omega \left( n+1/2\right )
\ee
The final analytical expression for both energy branches
\be
E_{\pm}=\frac{2kmn\omega}{2k^{2}-mn\omega}\pm\frac{2k^{2}\sqrt{m^{2}+2mn\omega-\frac{m^{3}n\omega}{2k^{2}}}}{2k^{2}-mn\omega}
\ee

\begin{figure*}
    \centering
    \includegraphics[scale=0.65]{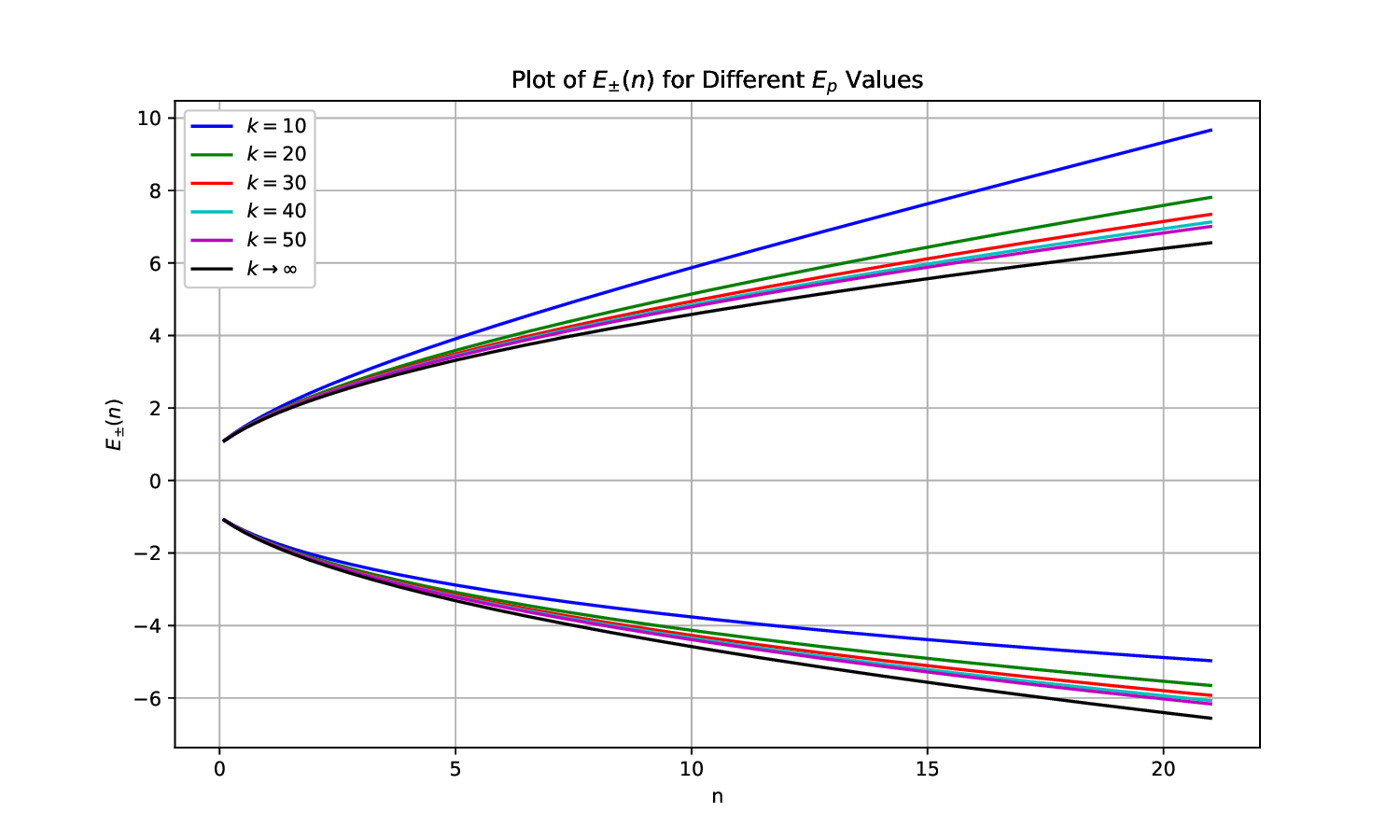}
    \caption{Plot of Energy Branches $E_{\pm}(n)$ for Different $k$ Values in Amelino-Camelia DSR}
    \label{fig2}
\end{figure*}

Figure \ref{fig2} illustrates the energy spectrum in this form.   This figure provides a detailed representation of the spectrum for the 1D Dirac Oscillator (DO) in the Amelino-Camelia DSR model, highlighting how the energy levels behave under the specific conditions considered.   It emphasizes the relationship between energy and quantum number $n$, as well as the impact of the deformation parameter $k$ on the structure of the spectrum.   By visually showing these elements, Figure 1 offers a comprehensive view of how the energy spectrum varies, allowing a deeper understanding of how the different parameters influence the overall energy profile within the context of the Amelino-Camelia model.
In the limit when $k\rightarrow \infty $, the  Result
\be
E_{\pm}(n)=\pm \sqrt{m^{2}+2nm\omega}
\ee
As in the previous case, the standard formula for the 1D Dirac oscillator is regained.

Finally, a notable observation can be made regarding the energy spectrum of the Amelino-Camelia DSR model as applied to the Dirac oscillator:

The Amelino-Camelia model exhibits a distinctive singularity in the oscillator energy spectrum. This singularity occurs at the critical deformation scale

$$
k_{c}(n) = \sqrt{\frac{m \omega n}{2}},
$$

where the energy diverges. This divergence arises mathematically from the modified energy expression: as the deformation parameter, linked to the Planck scale, approaches certain values, the denominator in the energy formula vanishes, causing the energy to blow up. Importantly, this divergence does not correspond to a physical breakdown or explosion of the oscillator but rather signals the breakdown of the low-energy approximation, indicating the growing influence of quantum gravity effects.

The position of the singularity explicitly depends on the quantum number $n$, which identifies the excitation level of the oscillator. As $n$ increases (representing higher-energy states), the singularity shifts to larger values of the deformation parameter. This means that more highly excited states require a stronger Planck-scale deformation before the energy spectrum becomes ill-defined. Conversely, for any fixed deformation parameter, there exists a maximum quantum number beyond which physical (finite) energy levels cannot exist, effectively imposing a natural cutoff in the excitation spectrum.

Thus, the presence of this singularity enforces an upper bound on the number of permissible oscillator excitations in strongly deformed regimes. Consequently, the AC-DSR model not only alters the energy levels of the Dirac oscillator, but also introduces intrinsic limits on the excitation spectrum, with the location of the singularity increasing with quantum number $n$.

It is worth emphasizing that this type of singularity does not appear in the MS (MS) DSR model, underscoring a fundamental difference between these two formulations in their treatment of Planck-scale modifications to quantum systems.

\section{ Some points about three dimensional case}

The examination of the Dirac oscillator under the principles of DSR — specifically via the Magueijo-Smolin and Amelino-Camelia models — uncovers substantial alterations to relativistic quantum dynamics at energy levels nearing the Planck scale.  Expanding these research from one-dimensional to three-dimensional (3D) Dirac oscillators constitutes a prospective frontier that would facilitate a more realistic and comprehensive understanding of relativistic quantum systems influenced by quantum gravity corrections.

 The 3D Dirac oscillator automatically encompasses more complex spin-orbit coupling and angular momentum configurations, rendering it an optimal option for investigating the nuanced Planck-scale deformations anticipated by DSR frameworks.  The deformation parameter $k$, regarded as the Planck energy scale, is anticipated to produce anisotropic and dimensionally dependent modifications to the energy spectrum, potentially resulting in changed degeneracies and adjusted transition rates.  These features are essential for connecting theory with future experimental or astrophysical discoveries.

 Furthermore, a 3D Dirac oscillator within the framework of DSR can provide insights into the interaction between quantum gravity effects and fundamental spinor fields in realistic spatial configurations.  The divergent responses of the MS and Amelino-Camelia models in one dimension, particularly the existence of spectrum singularities and excitation cutoffs in the latter, suggest that extending these models to three dimensions will enhance the understanding of the resilience and phenomenological characteristics of each framework.

 Future research on the 3D Dirac oscillator in DSR may concentrate on obtaining exact or approximate solutions, delineating the complete angular momentum spectrum, and examining dynamical features such as scattering and resonance phenomena in the context of Planck-scale deformations.  These developments will enhance our theoretical framework for quantum gravity phenomenology and may inform high-precision spectroscopic investigations designed to identify subtle deviations from standard relativistic quantum mechanics.

\section{Conclusion} \label{Conclusion}

DSR extends Einstein's special relativity by incorporating a universal length scale, such as the Planck scale, alongside the invariance of the speed of light. Although Einstein's theory focuses on the constancy of light speed, DSR modifies the space-time structure to account for quantum gravity effects, particularly at high energies or small scales near the Planck energy. This leads to modified Lorentz transformations and an energy-momentum relation that reflects these quantum gravitational effects, with noticeable deformations at high energies. Essentially, DSR adds quantum gravity considerations to special relativity, especially at the Planck scale. Within this context, the energy spectrum of the one-dimensional Dirac oscillator acquires corrections that depend on a deformation parameter $k$, typically identified with the Planck energy. Two prominent realisations of DSR, the MS and the Amelino-Camelia models, predict distinct patterns of deviation from the canonical relativistic oscillator. 
\begin{enumerate}
    \item for the case of the MS model: Within the MS realisation, the energy eigenvalues satisfy a transcendental equation containing terms proportional to $E$. Analytical and numerical analyses reveal that: 
\begin{itemize}
    \item  For small $k$, the level spacing deviates markedly from the standard $\sqrt{m^{2}+2nm\omega}$ form, leading to non-uniform shifts in both positive and negative branches.
    \item As $k$ increases, the modified spectrum gradually flattens toward the canonical result, indicating that DSR corrections become negligible atnck energies. 
    \item In the asymptotic $\lim k\to\infty$, one exactly recovers the relativistic Dirac oscillator spectrum: $E_{\pm}(n)\;=\;\pm\sqrt{m^{2}+2\,n\,m\,\omega}$. 
\end{itemize}
     \item Now, for the case of the Amelino-Camelia Model: The Amelino-Camelia construction leads to stronger modifications at lower values of $k$. Key observations include: 
\begin{itemize}
    \item At Planck-scale deformation, energy levels exhibit larger departures, with the gap between successive states becoming sensitive to n in a non-trivial way.
    \item Increasing k suppresses these anomalies, driving the spectrum smoothly back to its standard relativistic counterpart. 
    \item The convergence pattern toward $E_{\pm}(n)=\pm\sqrt{m^{2}+2nm\omega} $ is somewhat slower than in the MS model, reflecting the distinct functional form of the DSR deformation. 
\end{itemize}
\end{enumerate}
In conclusion, in both cases, the deformation effects decrease as $k$ increases, recovering the usual Dirac oscillator spectrum in $\lim k\to\infty$. 

These model calculations illustrate how DSR engenders finite-energy corrections to quantum systems as simple as the Dirac oscillator. The pronounced deviations at small k suggest possible phenomenological signatures in high-precision spectroscopic or astrophysical observations that probe energies approaching k. Moreover, comparing different DSR implementations provides insight into the robustness of Planck-scale predictions and guides experimental efforts aimed at detecting quantum-gravity effects in bound-state spectra.

\section{ Acknowledgment}
NJ has been funded by the Science Committee of the Ministry of Science and Higher Education of the Republic of Kazakhstan Program No. BR21881880.

\bibliography{main}

\end{document}